\documentclass[a4paper]{article}

\usepackage{multicol}
\usepackage{amsthm}
\usepackage{bm}

\usepackage[dvipdfmx]{graphicx}
\usepackage{bmpsize}
\usepackage{amssymb}
\usepackage{amsmath}
\usepackage{color}

\definecolor{dgreen}{rgb}{0.0, 0.5, 0.0}

\begin{document}

\fontsize{14pt}{16.5pt}\selectfont

\begin{center}
\bf{A simple model for ultradiscrete Hopf bifurcation}
\end{center}
\fontsize{12pt}{11pt}\selectfont
\begin{center}
Shousuke Ohmori$^{*)}$ and Yoshihiro Yamazaki\\ 
\end{center}

\noindent
\it{Department of Physics, Waseda University, Shinjuku, Tokyo 169-8555, Japan}\\

\noindent
*corresponding author: 42261timemachine@ruri.waseda.jp\\
~~\\
\rm
\fontsize{11pt}{14pt}\selectfont\noindent

\baselineskip 30pt

{\bf Abstract}\\
%\begin{abstract}

Dynamical properties of ultradiscrete Hopf bifurcation, 
similar to those of the standard Hopf bifurcation, 
are discussed by proposing a simple model 
of ultradiscrete equations with max-plus algebra.
In ultradiscrete Hopf bifurcation, 
limit cycles emerge depending 
on the value of a bifurcation parameter in the model.
The limit cycles are composed of a finite number 
of discrete states.
Furthermore, the model exhibits excitability.
The model is derived from two different dynamical models 
with Hopf bifurcation by means of ultradiscretization; 
it is a candidate for a normal form for 
ultradiscrete Hopf bifurcation.

\hrulefill
%
%
%
%\end{abstract}

%\noindent
%{\bf Key Words} : ultradiscretization, bifurcation, normal forms, discrete dynamical system\\

%\section{Introduction}
%\label{sec:1}

%
%%%%%%%%%%%%%%%%%%%%%%%%%%%%%%%%%%%%%%%%%%%%%%%%%%%%%%%%%%%%%%%%%%%%%%%%%%%%%%%%%%%%%%%%%%%%%%%%%%%%%%%%%%
%
%%%%%%%%%%%%%%%%%%%%%%%%%%%%%%%%%%%%%%%%%%%%%%%%%%%%%%%%%%%%%%%%%%%%%%%%%%%

% Bifurcation phenomena have been hugely studied 
% from viewpoint of nonlinear dynamical systems
% \cite{Prigogine,Guckenheimer,Strogatz,Nicolis,Murray, Robinson,Kuznetsov}.
% %
% As for Hopf bifurcation, Poincar\'{e}-Bendixson theorem 
% provides mathematical conditions of existence of limit cycles.
%
%Using these mathematical theorems we can not only assert that either specific bifurcations appear or disappear in a system but also obtain several mathematical properties for the bifurcations.
%
%Whereas, we often depend on numerical simulation analysis to obtain detail quantitative features of bifurcations. 
%
\ \\

For analysis of dynamical properties of nonlinear equations, 
piecewise linearization of them has often been carried out.
Especially, ultradiscretization, one of piecewise linearization methods, 
successfully retains and elucidates the essence of dynamical structures 
in nonlinear integrable systems 
\cite{Tokihiro1996,Grammaticos1997}.
The ultradiscretization method is summarized as follows.
Difference equations are derived 
from given continuous nonlinear equations.
By some limiting procedure, 
the difference equations are converted into 
other type of difference equations with max-plus algebra, 
which is called ultradiscrete equation.
This limiting procedure produces piecewise linearization 
of the original equations.
Note that the ultradiscretization can be applied 
to another type of dynamical systems 
such as non-integrable non-equilibrium dissipative systems 
and reaction-diffusion systems 
\cite{Nagatani1998,Murata2013,Ohmori2014,Matsuya2015,Murata2015,Gibo2015,Ohmori2016,Ohmori2020}.

Recently, we have applied ultradiscretization 
to bifurcation phenomena in one-dimensional dynamical systems\cite{Ohmori2020}.
Bifurcation phenomena have been hugely studied 
from viewpoint of continuous\cite{Guckenheimer,Strogatz,Nicolis} 
and discrete\cite{Robinson,Kuznetsov, Galor} dynamical systems.
In our study, focusing on the one-dimensional normal forms 
of saddle-node, transcritical, and pitchfork bifurcations, 
their ultradiscrete equations were derived 
and the dynamical properties 
for the obtained ultradiscrete equations were investigated.
In particular, we found that they possess 
``ultradiscrete bifurcations'', 
which has similar properties to the original bifurcations.
These ultradiscrete bifurcations can be visually understood 
by piecewise linear graphs of the ultradiscrete equations.
% 
%Also, we encountered another ultradiscrete bifurcation, similar to the flip bifurcation from supercritical bifurcation, where there is a stable cycle around a unstable fixed point.

%Note that since solutions of a ultradiscrete equation are formed with piecewise-linearity, 
%the features of the bifurcation occurring in a ultradiscrete equation can be easily grasped.
%
%In fact, for a ultradiscrete equation $U_{n+1}=\max(2U_n, C)$ derived from a normal form of a saddle-node bifurcation 

In this letter, we focus on Hopf bifurcation 
in two-dimensional dynamical systems.
First, we propose a model of ultradiscrete equations 
which exhibits a bifurcation similar to Hopf bifurcation.
Actually, our proposed model shows a dynamical transition 
between a monostable state and a state with limit cycles.
Next, we show that our model can be a normal form 
of ultradiscrete Hopf bifurcation 
by deriving it from two different nonlinear dynamical models.
%
%%%%%%%%%%%%%%%%%%%%%%%%%%%%%%%%%%%%%%%%%%%%%%%%%%%%%%%%%%%%%%%%%%%%%%%%%%%%

\begin{figure}[p]%[h!]
\begin{center}
\includegraphics[bb=0 0 960 720, width=6.5cm]{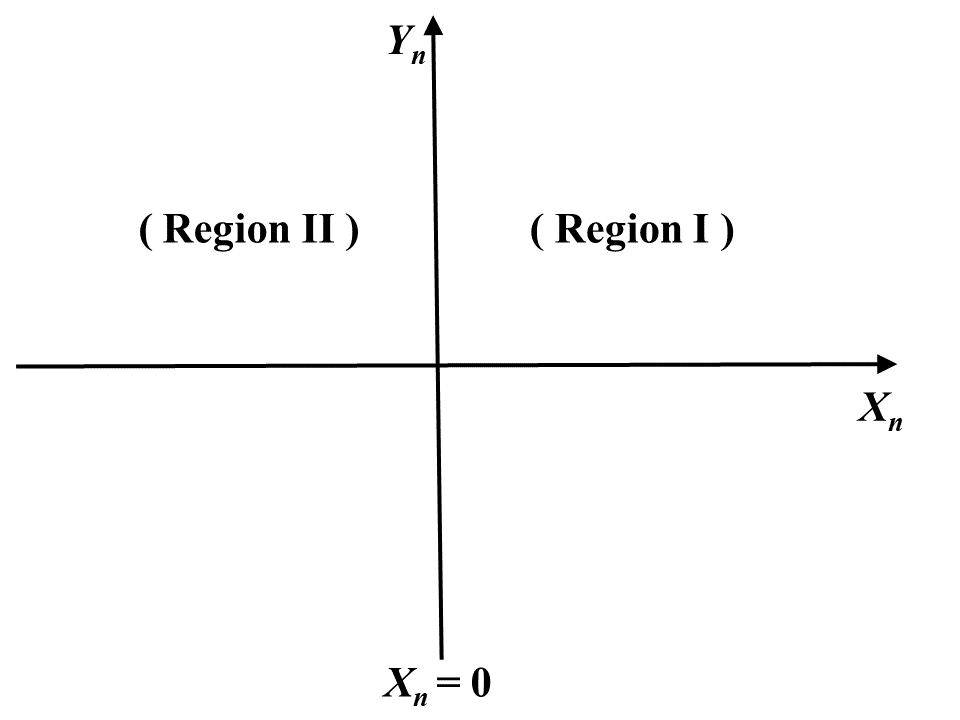}
\hspace{-20mm}
\includegraphics[bb=0 0 960 720, width=6.5cm]{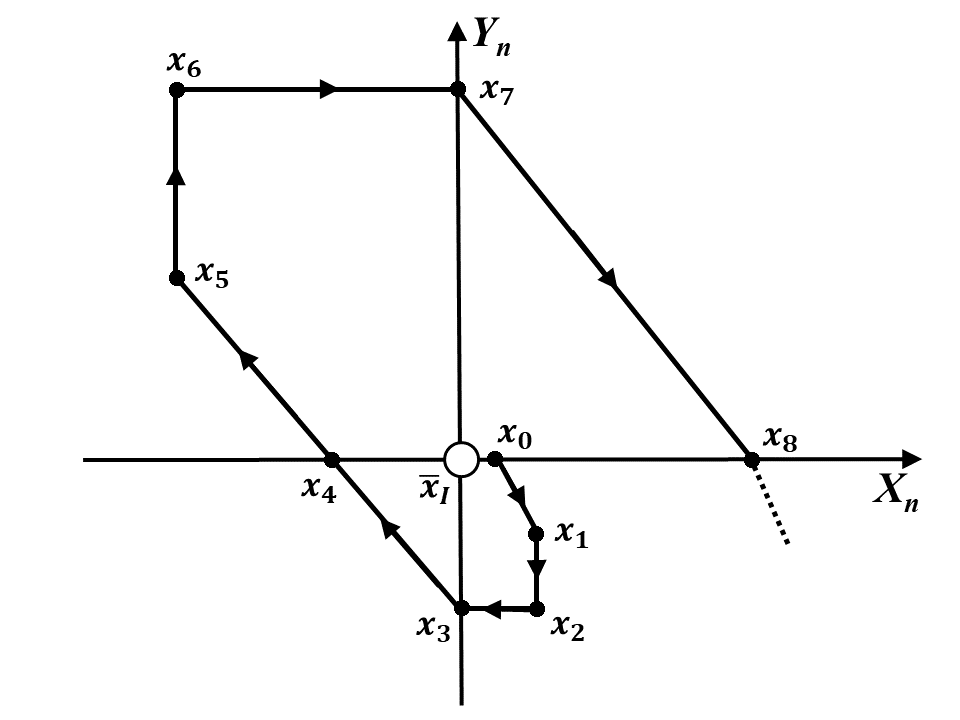}
\hspace{-20mm}
\includegraphics[bb=0 0 960 720, width=6.5cm]{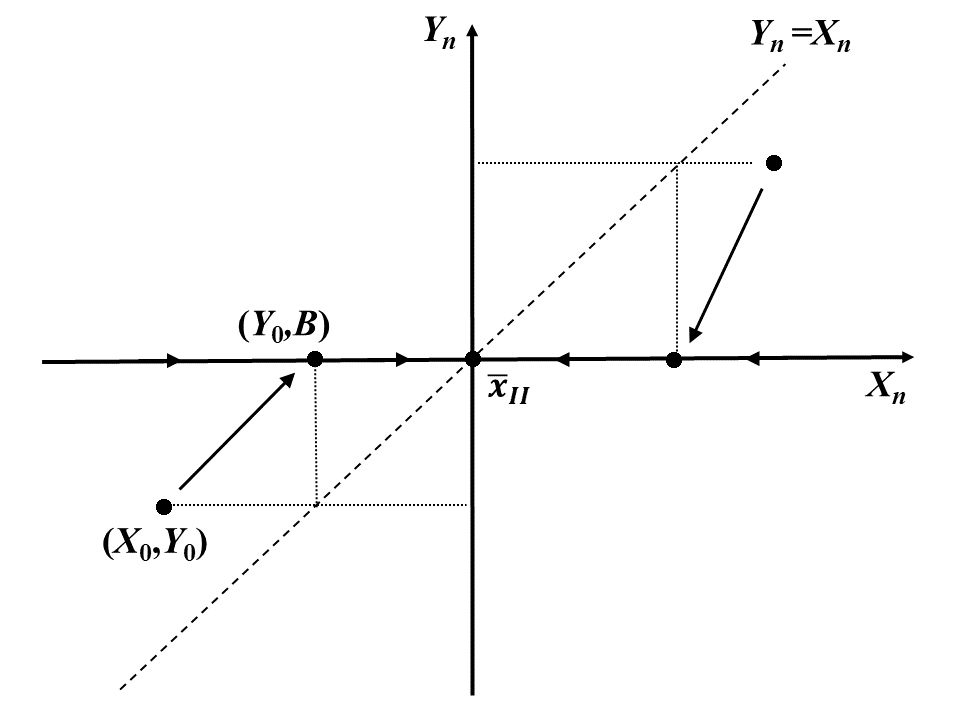}
\\
(a)
\hspace{4cm}
(b)
\hspace{4cm}
(c)\\
\caption{\label{Fig.1} 
  (a) Definition of regions I and II. 
  (b) A trajectory in the vicinity of the unstable focus $\bm{\bar x}_{I}$. 
  (c) Trajectories obtained from Eq. (\ref{eqn:2-2b}).}
\end{center}
\end{figure}

Let us focus on the following model 
consisting of max-plus equations with a bifurcation parameter $B$: 
\begin{eqnarray}
	X_{n+1} & = & Y_n + \max(0,2X_n),
	\label{eqn:2-1a} \\
	Y_{n+1} & = & B-\max(0,2X_n).
	\label{eqn:2-1b}
\end{eqnarray} 
%
%The flows of solutions for Eqs. (\ref{eqn:2-1a})-(\ref{eqn:2-1b}) can be visualized from viewpoint of dynamical systems.
%
Eqs. (\ref{eqn:2-1a})-(\ref{eqn:2-1b}) are considered 
as a discrete dynamical system $\bm{x}_{n+1}=\bm {F}(\bm{x}_{n})$ 
for the state variable $\bm{x}_n=(X_n,Y_n)$, equipping the evolution operator 
\begin{eqnarray}
	\bm {F} : R^2 \to R^2, (x,y) \mapsto (y+\max(0,2x), B-\max(0,2x)) .
	\label{eqn:2-1z}
\end{eqnarray}
A trajectory $\{\bm{x}_0, \bm{x}_1, \bm{x}_2, \dots \}
\left( \equiv \{ \bm{x}_n \} \right)$ 
from the initial point $\bm{x}_0 =(X_0,Y_0)$ 
is given by $\bm{x}_n=\underbrace{\bm {F}\circ\cdots\circ \bm {F}}_{n}(\bm{x}_0)=\bm {F}^n(\bm{x}_0) (n=1,2,\dots)$.
%
%Note that the right-hand sides of Eqs. (\ref{eqn:2-1a})-(\ref{eqn:2-1b}) have 
%the same second term, $\max(0,2X_n)$.
%
Here we set the following two regions I and II 
in $(X_n, Y_n)$ plane as shown in Fig.\ref{Fig.1} (a): 
(I) $X_n > 0$ and (II) $X_n\leq 0$.
In each region, Eqs. (\ref{eqn:2-1a})-(\ref{eqn:2-1b}) can be represented 
as the following matrix form.
\begin{description}
	\item[(Region I)] When $X_n > 0$, 
	Eqs. (\ref{eqn:2-1a})-(\ref{eqn:2-1b}) can be rewritten as %
	\begin{eqnarray}
		\left(
   			\begin{array}{ccc}
      		X_{n+1}  \\
      		Y_{n+1}  
   			\end{array}
  		\right)
		= 
		\left(
   			\begin{array}{ccc}
      		2 & 1  \\
      		-2 & 0  \\
   			\end{array}
  		\right)
		\left(
    		\begin{array}{ccc}
      		X_{n}  \\
      		Y_{n}  
    		\end{array}
  		\right)
		+
		\left(
   			\begin{array}{ccc}
      		0  \\
      		B  
  			\end{array}
  		\right) ,
		\label{eqn:2-2a}
	\end{eqnarray} 
	where Eq. (\ref{eqn:2-2a}) has the fixed point $\bm{\bar x}_I=(B,-B)$. 
	The matrix
	$
	\bm{A}_I = \left(
   			\begin{array}{ccc}
      		2 & 1  \\
      		-2 & 0  \\
   			\end{array}
  		\right)
	$
	satisfies Tr$\bm{A}_{I} = $ det$\bm{A}_I = 2$, where Tr and det stand for trace and determinant of a matrix, respectively.
	Therefore, the trajectory given by Eq. (\ref{eqn:2-2a}) is characterized 
	as a clockwise spiral source\cite{Galor}, 
	whose center is the unstable fixed point $\bm{\bar x}_I$.
	Fig. \ref{Fig.1}(b) shows the trajectory of $\{\bm{x}_{n}\}$  
	with $\bm{x}_0=(1,0)$ and $B=0$.
	
% when we set $B=0$, the spiral pattern shown by Fig. \ref{Fig.1} (b) is obtained from Eq. (\ref{eqn:2-2a}): 
% $\bm{x}_0=(1,0) \to 
% \bm{x}_{1}=(2,-2) \to 
% \bm{x}_{2}=(2,-4)\to 
% \bm{x}_{3}=(0,-4)\to 
% \bm{x}_{4}=(-4,0)\to 
% \bm{x}_{5}=(-8,8)\to
% \bm{x}_{6}=(-8,16)\to
% \bm{x}_{7}=(0,16)\to
% \bm{x}_{8}=(16,0)\to \cdots.$
%
%

	\item[(Region II)] 
When $X_n\leq 0$, 
the matrix form of Eqs. (\ref{eqn:2-1a})-(\ref{eqn:2-1b}) is 
	\begin{eqnarray}
		\left(
   			\begin{array}{ccc}
      		X_{n+1}  \\
      		Y_{n+1}  
   			\end{array}
  		\right)
		= 
		\left(
   			\begin{array}{ccc}
      		0 & 1  \\
      		0 & 0  \\
   			\end{array}
  		\right)
		\left(
    		\begin{array}{ccc}
      		X_{n}  \\
      		Y_{n}  
    		\end{array}
  		\right)
		+
		\left(
   			\begin{array}{ccc}
      		0  \\
      		B  
  			\end{array}
  		\right).
	\label{eqn:2-2b}
	\end{eqnarray} 
Equation (\ref{eqn:2-2b}) has the fixed point $\bm{\bar x}_{II}=(B,B)$.
For the matrix 
$\bm{A}_{II} = \left(
   		\begin{array}{ccc}
      0 & 1  \\
      0 & 0  \\
   		\end{array}
  	\right)$, 
	Tr$\bm{A}_{II} = $ det$\bm{A}_{II} = 0$.
The fixed point $\bm{\bar x}_{II}$ becomes a stable node.
Actually for any $\bm{x}_0=(X_0,Y_0)$, 
it is readily found that 
$\bm{x}_{1}=(Y_0,B)$ and $\bm{x}_{2}=(B,B)=\bm{\bar x}_{II}$;
Fig. \ref{Fig.1} (c) shows an example.
\end{description}
%

\begin{figure}[h!]
	\begin{center}
		\includegraphics[height=7.5cm]{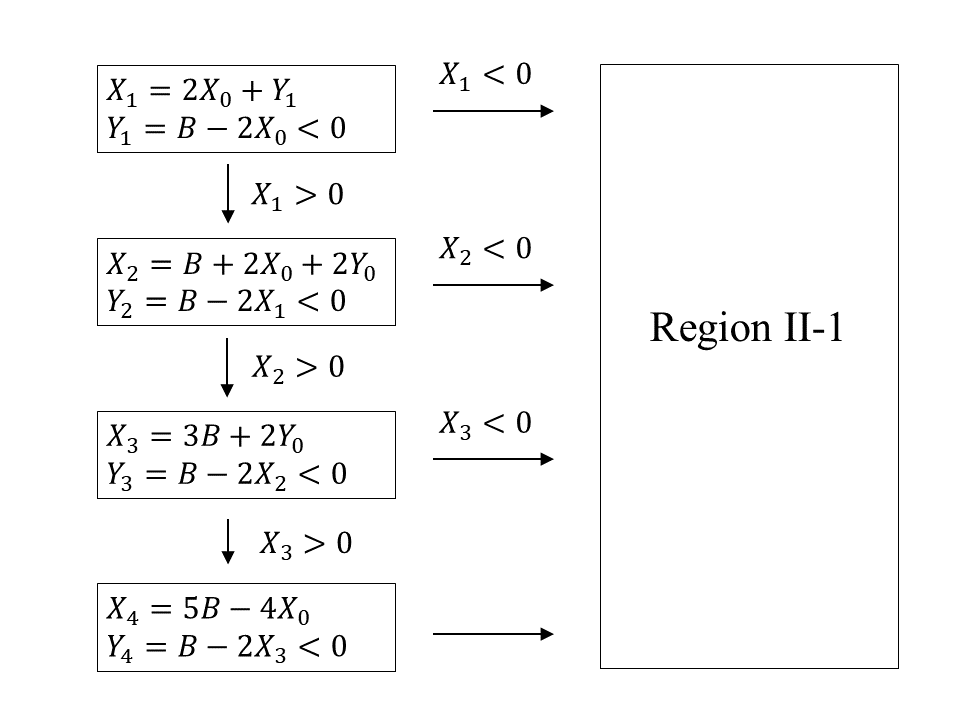}
		\caption{Flowchart for time evolutions of $\bm{x}_{n}$
		from the initial state $(X_0,Y_0)$ in region I. 
		$(X_0,Y_0)$ moves into region II-1 within four iteration steps.}
		\label{fig:flowchart}
	\end{center}
\end{figure}

\begin{figure}[h!]
\begin{center}
\includegraphics[width=7cm]{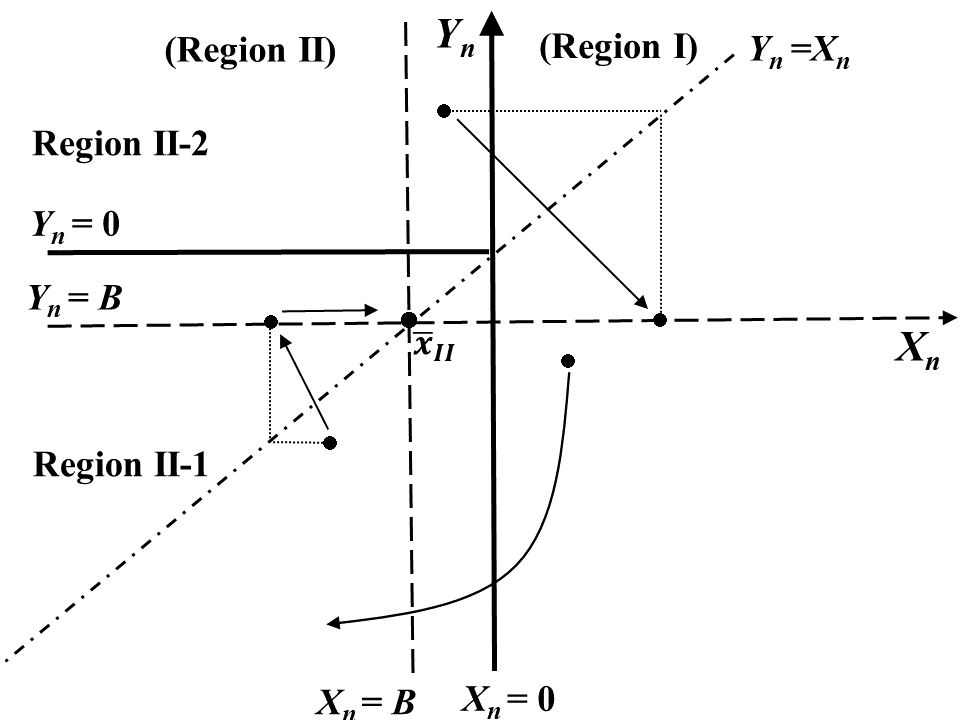}
\includegraphics[width=7cm]{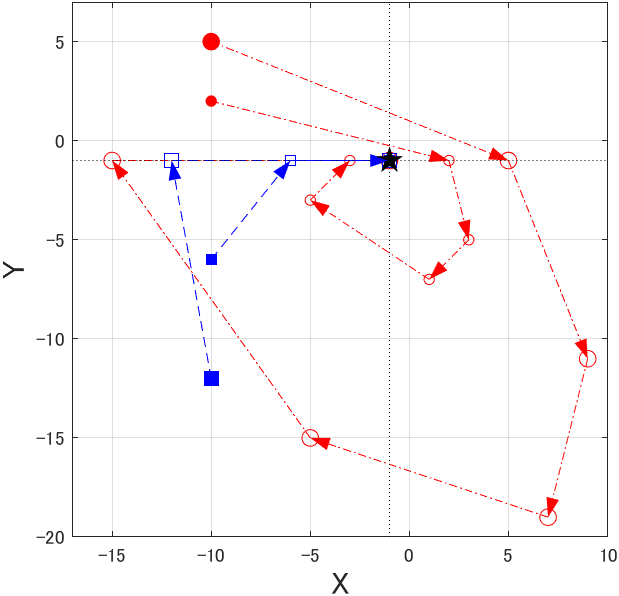}
\\
(a)
\hspace{7cm}
(b)\\
\caption{\label{Fig.2} 
	Trajectories obtained from Eqs. (\ref{eqn:2-1a})-(\ref{eqn:2-1b}) 
	with $B\leq 0$.
	(a) Schematic explanation.
	(b) Numerical results from four different initial states 
	described by filled circles and squares.
	Trajectories proceed in the direction of the arrows 
	and finally reach the stable fixed point $\bm{\bar x}_{II}$
	shown by the black star.}
\end{center}
\end{figure}

\begin{figure}[h!]
	\begin{center}
		\includegraphics[width=7cm]{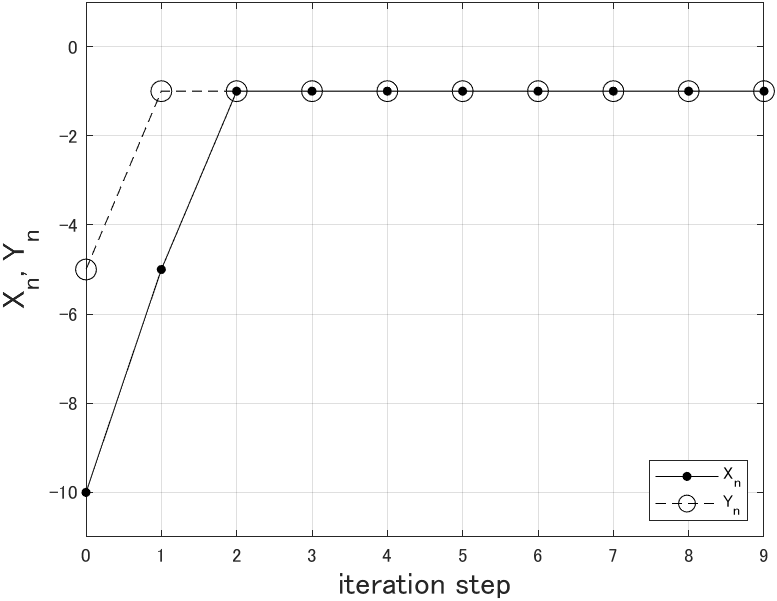}
		\includegraphics[width=7cm]{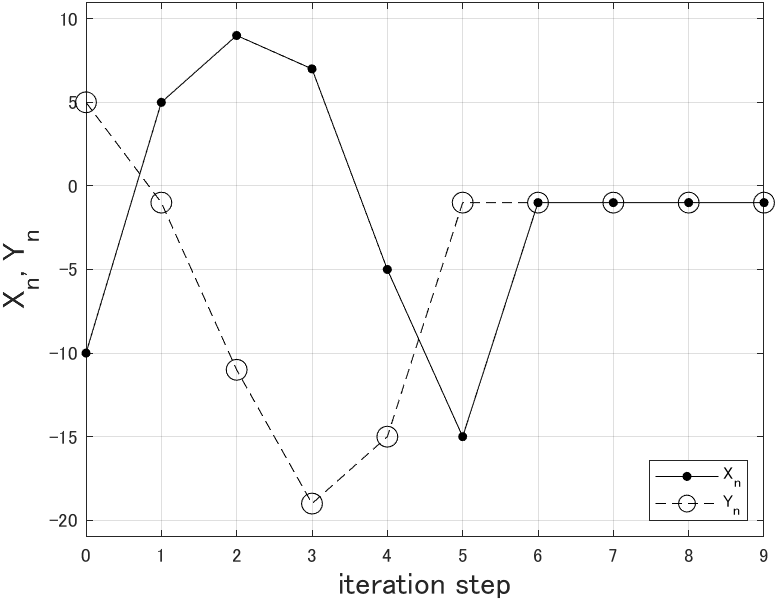}
	\\
	(a)
	\hspace{7cm}
	(b)\\
	\caption{\label{fig:excitable} 
	$(X_{n}, Y_{n})$ as a function of $n$ 
	for Eqs. (\ref{eqn:2-1a})-(\ref{eqn:2-1b}) with $B = -1$ 
	from two different initial conditions.}
	\end{center}
	\end{figure}

\begin{figure}[h!]
	\begin{center}
	\includegraphics[width=7cm]{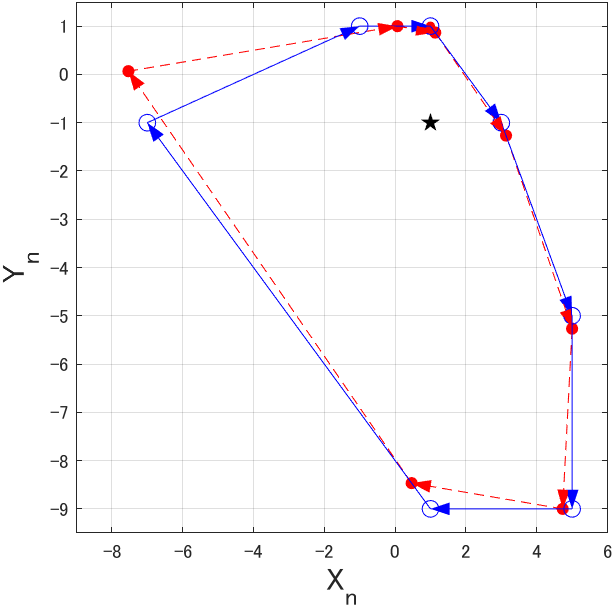}
	\includegraphics[width=7cm]{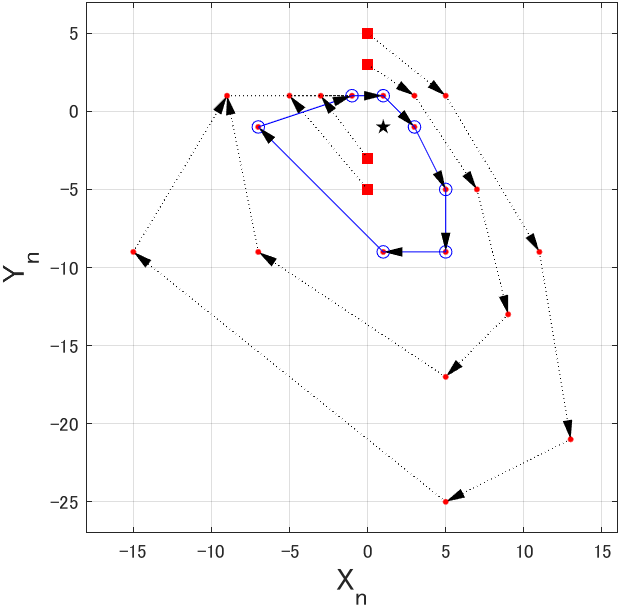}
	\\
	(a)
	\hspace{7cm}
	(b)\\
	\caption{\label{fig:lc} 
	  (a) The two limit cycles $\mathcal{C}$ (open circles) 
		and $\mathcal{C}_s$ (filled circles).
		(b) Examples of trajectories starting 
		from four different filled squares.
		The trajectories finally converge into $\mathcal{C}$.
		The star in each figure shows $\bm{\bar x}_{I}$.}
	\end{center}
	\end{figure}

%%%%%%%%%%%%%%%%%%%%%%%%%%%%%%%%%%%%%%%%%%%%%%%%%%%%%%%%%%%%%%%%%%%%%%%%%%%%
Taking these dynamical properties in regions I and II into account, 
bifurcation for Eqs. (\ref{eqn:2-1a})-(\ref{eqn:2-1b}), 
can be grasped as follows.
(i) When $B \leq 0$, both $\bm{\bar x}_{I}$ and $\bm{\bar x}_{II}$ 
are in region II; 
$\bm{\bar x}_{II}$ becomes a fixed point, but $\bm{\bar x}_{I}$ does not.
Then, Eqs. (\ref{eqn:2-1a})-(\ref{eqn:2-1b}) have a unique fixed point $\bm{\bar x}_{II}=(B,B)$.
(i-a) When $\bm{x}_{0}=(X_0,Y_0)$ belongs to ``region II-1'', 
which means $X_0 \leq 0$ and $Y_0 \leq 0$, 
$\bm{x}_{1}=(Y_0,B)$ and $\bm{x}_{2}=(B,B)=\bm{\bar x}_{II}$. 
%
%Call this subregion $\{\bm{x}_{n}=(X_n,Y_n), X_n \leq 0, Y_n \leq 0\}$ in Region II ``Region II-1''.
%
Then any initial point in region II-1 reaches 
$\bm{\bar x}_{II}$ with two iteration steps. 
(i-b) $\bm{x}_{0}$ in region I moves into region II-1
within four iteration steps as shown in Fig.\ref{fig:flowchart}.
Then, any initial point in region II-1 reaches $\bm{\bar x}_{II}$ 
within six steps. 
%
% In particular, the following characteristic properties can be verified for $\bm{x}_{0}=(X_0,B)$ with $X_0 >0$ ;
% %
% %\begin{description}
% 	%\item[(i)] 
% 		When $0 < X_0 \leq \frac{-B}{4}$, 
% 		$\bm{x}_{1}=(2X_0+B,-2X_0+B)$ is contained in Region II-1.
% %
% 	%\item[(ii)] 
% 		When $\frac{-B}{4} < X_0 \leq \frac{-5B}{4}$, $\bm{x}_{2}=(2X_0+3B,-4X_0-B)$ is contained in Region II-1.
% % 
% 	%\item[(iii)] 
% 		When $\frac{-5B}{4} < X_0 $, $\bm{x}_{3}=(5B,-4X_0-5B)$ is contained in Region II-1.
% %\end{description}
% %
% Therefore, $\bm{x}_{0}=(X_0,B)$ with $X_0 >0$ converges to $\bm{\bar x}_{II}$ at most five iteration steps.
% %
(i-c) When $\bm{x}_{0}$ is in ``region II-2'', 
which shows $X_0 \leq 0$ and $Y_0 > 0$, 
$\bm{x}_{1}=(Y_0,B)$ belongs to region I.
Therefore, $\bm{x}_{0}$ reaches $\bm{\bar x}_{II}$ 
within seven steps from the property (i-b).
From (i-a)-(i-c), it is found that the fixed point 
$\bm{\bar x}_{II}=(B,B)$ is stable.
The trajectories of $\{ \bm{x}_{n} \}$ in this case 
are shown in Fig. \ref{Fig.2}.
Figure \ref{Fig.2} also shows that excitability occurs 
when $\bm{x}_{0}$ is in region II-2.
Figure \ref{fig:excitable} shows time evolutions of 
$(X_{n}, Y_{n})$ from two different initial conditions.
By comparing with Fig.\ref{fig:excitable}(a), 
it is found that excitability occurs 
in Fig.\ref{fig:excitable}(b).
(ii) When $B>0$, $\bm{\bar x}_{I}=(B,-B)$ becomes a unique unstable fixed point.
We find that there exist only two different clockwise periodic solutions, 
$\mathcal{C}$ and $\mathcal{C}_s$ as shown in Fig.\ref{fig:lc}(a), 
which are composed of the following seven points, respectively: 
$\mathcal{C} = \{(B,B) \rightarrow (3B,-B) 
\rightarrow (5B,-5B) \rightarrow (5B,-9B) 
\rightarrow (B,-9B) \rightarrow (-7B,-B) 
\rightarrow (-B,B) \left[ \rightarrow (B,B) \right] \}$, 
$\mathcal{C}_s = \{ \left(\frac{B}{15},B \right) 
\rightarrow \left(\frac{17B}{15}, \frac{13B}{15} \right)
\rightarrow \left(\frac{47B}{15}, -\frac{19B}{15} \right) 
\rightarrow \left(5B, -\frac{79B}{15} \right) 
\rightarrow \left(\frac{71B}{15}, -9B \right) 
\rightarrow \left(\frac{7B}{15}, -\frac{127B}{15} \right) 
\rightarrow \left(-\frac{113B}{15}, \frac{1B}{15} \right) 
\left[ \rightarrow \left(\frac{B}{15}, B \right) \right] \}$.
%
% \color{red}
% それぞれのサイクルのベイスンは？
% 他にサイクルが存在する可能性はない？
% \color{black}
%
% Two different clockwise periodic solutions $\mathcal{C}$ and $\mathcal{C}_s$ 
% with seven period emerge
% on a piecewise linear closed line $\Omega$ around $\bm{\bar x}_{I}$.
%
It is also found that a trajectory with any initial condition 
except for $\bm{\bar x}_{I}$ is finally absorbed 
in $\mathcal{C}$ or $\mathcal{C}_s$.
Therefore, $\mathcal{C}$ and $\mathcal{C}_s$ are considered 
as limit cycles.
Figure \ref{fig:lc}(b) shows trajectories 
from four different initial conditions; 
they finally converge into $\mathcal{C}$.

Here we discuss basins for $\mathcal{C}$ and $\mathcal{C}_s$.
It is confirmed that any trajectory has a point 
with $Y_{n} = B$ at a certain iteration step $n$.
In other words, a point on 
the line $l_{B}$, $\bm{x}_n = (X_n,B)$  
where $X_n \in (-\infty, \infty)$, exists in all trajectories.
%
% Now we consider the subsets $J_{+}$ and $J_{-}$ on $l_{B}$, 
% where $X_n < 0$ in $J_{-}$ and $X_n \leq 0$ in $J_{+}$.
% %
% $\bm{x}_0$ in region II is evolved to 
% a point in $J_{+}$ within two iteration steps.
% %
% Actually, $\bm{x}_2 =(B,B) \in \mathcal{C}$ 
% when $\bm{x}_0$ is in region II-1, 
% and $\bm{x}_1 \in J_{+}$ when $\bm{x}_0$ is in region II-2.
% %
% Furthermore, any point in region I moves spirally 
% and goes into region II based on Eq. (\ref{eqn:2-2a}). 
%
It is found that all trajectories with a point on $l_{B}$ 
except for $\left( \frac{B}{15},B \right)$ 
and $\left( \frac{17B}{15},B \right)$
are finally absorbed into $\mathcal{C}$.
And any trajectory with $\left( \frac{B}{15},B \right)$ 
or $\left( \frac{17B}{15},B \right)$ finally goes into $\mathcal{C}_s$.
Therefore, there exist only two limit cycles 
$\mathcal{C}$ and $\mathcal{C}_s$ in this model.
%
%(The detailed discussion and explanation 
%will be reported elsewhere.)
%
%
% \begin{figure}[h!]
% \begin{center}
% %\includegraphics[width=7cm]{Fig2a_210205.png}
% \includegraphics[width=7cm]{Fig2l_210205.png}
% \includegraphics[width=7cm]{Fig2b_210205.png}
% \\
% (a)
% \hspace{4cm}
% (b)\\
% \caption{\label{Fig.3} 
% 	%the sketches of (a) flows of solutions for Eqs. (\ref{eqn:2-1a})-(\ref{eqn:2-1b}) with $B\leq 0$, 
% 	The sketches of (a) the subregions $J_0,J_1,J_2,J_3$ in the line $l_{B}$,  
% 	(b) the piecewise linear closed line $\Omega$ and periodic solutions $\mathcal{C}$ and $\mathcal{C}_s$ for $B>0$.
% 	The dots lines show $\Omega$ constructed as the union of $I_i$, and
% 	the filled circles and triangles are points composing $\mathcal{C}$ and $\mathcal{C}_s$, respectively.}
% \end{center}
% \end{figure}
% %

%%%%%%%%%%%%%%%%%%%%%%%%%%%%%%%%%%%%%%%%%%%%%%%%%%%%%%%%%%%%%%%%%%%%%%%%%%%%%%%%%%%%%%%%%%%%%%%%%%%%%
Now we show that Eqs. (\ref{eqn:2-1a})-(\ref{eqn:2-1b}) become 
a candidate of a normal form of ultradiscrete Hopf bifurcation.
Through ultradiscretization, 
Eqs. (\ref{eqn:2-1a})-(\ref{eqn:2-1b}) can be derived from 
the following two different dynamical systems described by partial differential equations.
\ \\
(i) Sel$^{\prime}$kov model\cite{Strogatz,Selkov1968}
\begin{eqnarray}{}
  		\displaystyle\frac{d x}{d t}
    		& = & -x+ay+x^2y, 
  		\label{eqn:3-1a}
  \\
 	 \displaystyle\frac{d y}{d t}
    & = &  b-ay-x^2y, 
		\label{eqn:3-1b}
\end{eqnarray}
where $a$ and $b$ are positive.
They can be considered as bifurcation parameters 
for Hopf bifurcation.
Actually, eqs. (\ref{eqn:3-1a})-(\ref{eqn:3-1b}) exhibit 
Hopf bifurcation when $a$ and $b$ satisfy 
$b^2 = \frac{1}{2}(1-2a\pm \sqrt{1-8a})$.
Ultradiscretiazation of Sel$^{\prime}$kov model can be performed 
in the following way.
By tropical discretization\cite{Murata2013}, 
the following difference equations are adopted for 
eqs. (\ref{eqn:3-1a})-(\ref{eqn:3-1b}), 
	\begin{eqnarray}{}
  				x_{n+1} = \frac{x_n+\Delta t(ay_n+x^2_ny_n)}{1+\Delta t}, 
   				\label{eqn:3-2a}
  		\\
 				y_{n+1} = \frac{y_n+\Delta t b}{1+\Delta t(a+x_n^2)}, 
   				\label{eqn:3-2b}
 	\end{eqnarray}
where $\Delta t$ is the discretized time interval.
$x_n=x(n\Delta t)$, $y_n=y(n\Delta t)$, where $n$ is positive integer.
%
%
%Such a replacement makes Eq.(\ref{eqn:4}) possible to be symmetrized 
%by the appropriate transformations.)
%
The variable transformations, 
\begin{eqnarray}
	\begin{cases}
		\Delta t = e^{T/\varepsilon}, \;\;\;
		x_n = e^{X_n/\varepsilon}, \;\;\;
		y_{n}= e^{Y_{n}/\varepsilon }, \\
		a  = e^{A/\varepsilon}, \;\;\;
		b =e^{B/\varepsilon }, \;\;\;
		\end{cases}
	\label{eqn:3-3}
\end{eqnarray}
are applied to eqs.(\ref{eqn:3-2a})-(\ref{eqn:3-2b}), 
and the ultradiscrete limits
	\begin{eqnarray}
\begin{cases}
		\displaystyle\lim_{\varepsilon  \to +0} \varepsilon  \log(e^{A/\varepsilon }+e^{B/\varepsilon }+\cdot \cdot \cdot )~=~\max(A,B,\dots),\\
		\displaystyle\lim_{\varepsilon  \to +0} \varepsilon  \log(e^{A/\varepsilon }\cdot e^{B/\varepsilon }\cdot \dots ) = A+B+ \dots.
\end{cases}
		\label{eqn:0}
	\end{eqnarray}
are performed. 
Then, the ultradiscrete equations for Sel$^{\prime}$kov model 
are obtained as
	\begin{eqnarray}
		X_{n+1} & = & \max(X_n, T+\max(A+Y_n,2X_n+Y_n))-\max (0,T),
		\label{eqn:3-4a} \\
		Y_{n+1} & = & \max(Y_n,T+B)-\max(0,T+\max(A,2X_n)).
		\label{eqn:3-4b}
	\end{eqnarray} 
Assuming that $T \geq \max\{0,-A, Y_n-B, -(X_n+Y_n)\}$ 
for all $n$ and $A=0$,
eqs. (\ref{eqn:3-4a})-(\ref{eqn:3-4b}) are identical 
to eqs.(\ref{eqn:2-1a})-(\ref{eqn:2-1b}). 
\ \\
(ii) Lengyel model  
%for the chlorine dioxide-iodine-malonic acid reaction 
\cite{Lengyel1990, Lengyel1991}, 
\begin{eqnarray}{}
  		\displaystyle\frac{d x}{d t}
    		& = & r -x-\frac{4xy}{1+x^2}, 
  		\label{eqn:4-1a}
  \\
 	 \displaystyle\frac{d y}{d t}
    & = &  x(1-\frac{s y}{1+x^2}), 
		\label{eqn:4-1b}
\end{eqnarray}
where $r$ and $s$ are positive bifurcation parameters 
for Hopf bifurcation.
Ultradiscrete equations of Eqs. (\ref{eqn:4-1a})-(\ref{eqn:4-1b}) 
are obtained in the similar way to the case 
of Eqs. (\ref{eqn:3-1a})-(\ref{eqn:3-1b}). 
By tropical discretization, 
the following difference equations are derived 
from Eqs. (\ref{eqn:4-1a})-(\ref{eqn:4-1b}), 
\begin{eqnarray}{}
  x_{n+1} = \frac{x_n+\Delta t(r -x_n)}{1+\Delta t(\frac{4y_n}{1+x^2_n})}, 
  \label{eqn:4-2a}\\
  y_{n+1} = \frac{y_n+\Delta t x_n}{1+\Delta t(\frac{s x_n}{1+x_n^2})}.
	\label{eqn:4-2b}
\end{eqnarray}
The variable transformations, 
\begin{eqnarray}
	\begin{cases}
		\Delta t = e^{T/\varepsilon}, \;\;\;
		1 - \Delta t = e^{M/\varepsilon}, \;\;\;
		x_n = e^{X_n/\varepsilon}, \;\;\;
		y_{n}= e^{Y_{n}/\varepsilon }, \\
		r  = e^{R/\varepsilon}, \;\;\;
		s =e^{S/\varepsilon }, \;\;\;
		\end{cases}
	\label{eqn:4-3}
\end{eqnarray}
and ultradiscrete limits (\ref{eqn:0}) are carried out.
Then, the ultradiscrete equations for Lengyel model are obtained as
	\begin{eqnarray}
		X_{n+1} & = & R-Y_n+\max(0,2X_n),
		\label{eqn:4-4a} \\
		Y_{n+1} & = & -S+\max(0,2X_n),
		\label{eqn:4-4b}
	\end{eqnarray} 
on the assumption $T \geq \max\{0,X_n+M-R, Y_n-X_n, |X_n|-S\}$ for all $n$.
It is noted that Eqs.(\ref{eqn:4-4a})-(\ref{eqn:4-4b}) 
have counterclockwise limit cycles 
reflected on the signs in the equations.
Therefore taking the variable transformations $Y_n \to -Y_n$ 
and setting $R=0, S= B$ in Eqs.(\ref{eqn:4-4a})-(\ref{eqn:4-4b}), 
Eqs.(\ref{eqn:2-1a})-(\ref{eqn:2-1b}) can be reproduced.
%
%Therefore, we can exhibit a novel relationship between different nonlinear systems 
%(\ref{eqn:3-1a})-(\ref{eqn:3-1b}) and (\ref{eqn:4-1a})-(\ref{eqn:4-2a}) 
%that both models have the same ultradiscrete dynamical system Eqs. (\ref{eqn:2-1a})-(\ref{eqn:2-1b}).
%
%$Y_n \to -Y_n$?ｽﾉゑｿｽ?ｽ?ｽChemical model?ｽﾌ抵ｿｽ?ｽ?ｽ?ｽU?ｽ?ｽ?ｽ?ｽ?ｽ?ｽ?ｽﾌ趣ｿｽ?ｽ?ｽ?ｽ?ｽ?ｽt?ｽﾌ話
%Note that the dynamical properties for Eqs.(\ref{eqn:4-6})-(\ref{eqn:4-7}) are inherited from Eqs.(\ref{eqn:2-8})-(\ref{eqn:2-9}) 
%under the transformations (\ref{eqn:4-8}).
%
%Note that the orientation of iteration steps for $\mathcal{C}'$ ($\mathcal{C}'_s$) 
%are opposite to of $\mathcal{C}$ ($\mathcal{C}_s$) because of the transformation $Y_{n} \to -Y_{n}$.
%
%It seems that this dynamical difference is reflected on the difference of the dynamical properties of the original models.
%
%Actually, the orientation of rotation for the limit cycle described in Sel$^{\prime}$kov model (\ref{eqn:2-1})-(\ref{eqn:2-2}) 
%is opposite to that for the oscillating chemical reaction model (\ref{eqn:4-1})-(\ref{eqn:4-2}).
%

In conclusion, we have proposed the simple model 
for ultradiscrete Hopf bifurcation.
Depending on the value of the bifurcation parameter, 
the model exhibits excitability 
and possesses the limit cycles.
The model is derived 
from the two different nonlinear dynamical models, 
Sel$^{\prime}$kov model and Lengyel model; 
it can be a normal form of ultradiscrete Hopf bifurcation.
Further investigation 
for basins of $\mathcal{C}$ and $\mathcal{C}_s$ 
and characterization for their periodicity 
will be future problems.

% \txred{
% あとは問題点を挙げる to be tackled in the future.
% }
% \color{blue}
% ↓以下、暫定です。
% Some questions that   
% what factor determines the period of $\mathcal{C}$ and $\mathcal{C}_s$,
% can the basin of $\mathcal{C}_s$ in Region I whose point is not absorbed in $\mathcal{C}$ be visualized, 
% and what spatiotemporal dynamics occur from the excitable or the limit cycle system of the present model with spatial diffusion effects   
% will be tackled in the future.
% \color{black}

%
%%%%%%%%%%%%%%%%%%%%%%%%%%%%%%%%%%%%%%%%%%%%%%%%%%%%%%%%%%%%%%%%%%%%%%%%%%%%%%%%%%%%%%%%%%%%%%%%%%%%%%%%%

\bigskip

\noindent
{\bf Acknowledgement}

The authors are grateful to 
Prof. M. Murata, at Tokyo University of Agriculture and Technology, 
Prof. K. Matsuya at Musashino University,
Prof. D. Takahashi, 
Prof. T. Yamamoto, and Prof. Emeritus A. Kitada 
at Waseda University for useful comments and encouragements. 
%
%Also, we greatly appreciate the valuable comments from an reviewer.
This work was
supported by Sumitomo Foundation, Grant Number 200146.

\bigskip

%\noindent
%{\bf Data Availability}

%Data sharing is not applicable to this article as no new data were created or analyzed in this study.\\

\end{document}